# CP Properties of Leptons in Mirror Mechanism

Igor T. Dyatlov *
*Scientific Research Center "Kurchatov Institute"*
*Petersburg Institute of Nuclear Physics, Gatchina, Russia*

Formation of quark and lepton mass matrices through intermediate states of heavy mirror fermions is capable of reproducing main qualitative properties of weak mixing matrices (CKM and PMNS matrices). The reproduction includes the hierarchy of CKM matrix elements, a general form of the PMNS matrix, involving the smallness of the neutrino mixing angle $\theta_{13}$, and leads to very small neutrino masses. For leptons, these properties appear only for the Dirac nature of SM neutrino and its inverse spectrum. In such a lepton system, not only do spontaneous mirror symmetry violation and the observed mass hierarchy of charged leptons $(e, \mu, \tau)$ govern the structure of the PMNS matrix but also permit assessment of its allowed complexity, that is, the CP properties of leptons. The PMNS matrix then contains no Majorana phases and its Dirac phase $\delta_{CP}$ corresponds to $|sin\ \delta_{CP}|$, which is significantly smaller than unity.

## 1. Introduction

The potential complexity of the lepton weak mixing matrix (WMM)—the source of CP violation in lepton systems—is of particular interest in connection with the evolving conception of leptogenesis as the cause of baryon asymmetry in the Universe ([1], review [2]).

Preliminary data obtained by T2K Group [3] indicate that the Dirac phase of lepton CP violation has a large value: $\delta_{CP} \simeq -\pi/2$, $|sin\delta_{CP}| \approx 1$. This result and its interpretation, however, are not final [4]; six experimental groups are proposing to continue work on $\delta_{CP}$ determination [2,5].

In earlier papers, the author proposed a mechanism of mirror symmetry (MS) violation to describe a number of Standard Model (SM) properties that can be explained phenomenologically but are difficult to understand theoretically. The main objective of [6-9] is to reproduce WMM structures for quarks and leptons. This is the guiding principle for selecting the system and phenomena that can reproduce the observed picture.

The spontaneous violation of SM's mirror symmetry generalization [6-9] appears to provide a mechanism whereby the qualities of the Cabbibo-Kobayashi-Maskawa (SKM) matrix—that is, the hierarchy of its elements [10]—can easily be achieved. This hierarchy appears to be directly related with the hierarchical character of quark mass spectra. Furthermore, qualitative properties of the Pontecorvo-Maki-Nakagawa-Sakata (PNMS) matrix [11], which do not obey any distinct pattern, are fully reproduced in the mirror scenario, including even such details as the smallness of the neutrino mixing angle $\theta_{13}$ [12]. In [8, 9], it is determined that the PMNS matrix can be easily reproduced only for the Dirac nature and inverse spectrum of three generations of neutrinos. In such a system, the proposed analog of the see-saw mechanism (see review [13]) leads to a mass formula that demonstrates a possibility of very small neutrino masses [11] even with a higher certitude than the see-saw mechanism.

*E-mail: dyatlov@thd.pnpi.spb.ru

The appearance of the Dirac fermion in the see-saw mechanism is not a trivial problem. The see-saw is characterized by the presence of Majorana terms that do not conserve lepton numbers, which typically leads to Majorana particles. In the spontaneously violated MS, we can construct the Dirac scenario that meets the observed properties. It appears that this rare situation would require a very restrictive selection of ratios of system parameters. However, based on the results obtained for this scenario—indicative of a large difference between neutrino and charged lepton masses and affinity to the PMNS matrix—the character of the PMNS matrix complexity warrants here close attention. These properties are discussed in this paper.

A direct calculation of the Dirac phase $\delta_{CP}$ in the MS mechanism is beyond the limits of approximations used in this paper. Such calculation is not considered reasonable due to the abundance of parameters defining this phase in the most general possible case of the chosen scheme. It will be shown, however, that the PMNS matrix complexity in the proposed scenario is always accompanied by the small factor $\sim m_e / m_\mu \approx 0.005$ (masses of $e$-, $\mu$-leptons). Therefore, the complex part of the element $V_{3e}$ in its standard form [11]

$$V_{3e} = \sin\theta_{13}\, e^{-i\delta_{CP}}, \tag{1}$$

has substantial reasons to be less than its observed absolute value $|sin\,\theta_{13}| \approx 0.14\text{-}0.16$, which means a distinct smallness

$$|\sin\delta_{CP}| < 1 \tag{2}$$

and does not meet the expectations in [3] (Appendix 2).

Furthermore, there are no Majorana phases in the MS scenario being discussed (see Section 2) and for approximations used in the lepton WMM.

This paper considers the lowest approximation of the ratio of SM fermion masses to the heavy masses of mirror analogs: $m_{SM} / M_{mir} \ll 1$. The enormous value of $M_{mir}$ is not just a convenient choice for the scenario under consideration but a necessary condition for the reproduction of the observed properties. In the MS scenario being discussed, all mirror fermions have very large masses.

In Section 2, conditions for the appearance of Dirac neutrinos are determined for the MS analog of the see-saw mechanism. In Section 3, a mass matrix is derived for such neutrinos, and their wavefunctions in the space of generation indices are determined. The complex parameters of the PMNS matrix are discussed in Section 4. Appendix 1 shows that transitions between different representations of neutral leptons used in this paper do not affect the kinetic parts of the Lagrangian of the system but lead to a 100% violation of the lepton numbers in processes with heavy mirror neutrinos. In Appendix 2, a formula for $V_{3e}$ complexity (1) is written out for the parameters of the proposed MS model.

## 2. Dirac Neutrino in MS See-Saw Mechanism

The observed differences in the properties of charged lepton and neutrino spectra and quark and lepton WMMs [11] can be explained by the presence in the neutrino part of the Lagrangian of Majorana terms that do not conserve lepton numbers. This, generally speaking, results in these neutrinos being Majorana as well. In this situation, the appearance of Dirac particles becomes possible if Majorana neutrinos are present as pairs with equal (in absolute value) masses. Two Majorana states with equal masses form one Dirac state. This condition imposes restrictions on the structure of the part of the Lagrangian responsible for particle masses. For the MS model, such restrictions also include complex properties.

In [6-9], MS fermions are expressed in terms of Dirac operators ($\Psi_a^{(f)}$)

$$\Psi_{LR} = \psi_L + \Psi_R \left(T_W = \frac{1}{2}\right), \quad \Psi_{RL} = \psi_R + \Psi_L \ (T_W = 0) \tag{3}$$

for three generations $a, b = 1,2,3$ and two flavors $f = \bar{u}, \bar{d}$. In [3], *L*, *R* are chiralities, $T_W$ is the weak isospin of the *SU*(2) group. A system with fermion states (3) is apparently invariant to the replacement:

$$L \leftrightarrow R, \quad \Psi \leftrightarrow \psi. \tag{4}$$

"Mirror symmetry" violation in (4) is the transition from the $\Psi_{LR}$ and $\Psi_{RL}$ states to the $\Psi$ and $\psi$ ones with different masses. The violation mechanism is constructed by analogy with SM using suitable scalars. To separate $\Psi$ from $\psi$, each scalar used should be matched by a similar pseudoscalar [7].

Due to their chiral properties, kinetic and all gauge couplings automatically separate $\Psi$ from $\psi$:

$$\mathcal{L}_{SM}(\Psi_{RL}, \Psi_{LR}) \equiv \mathcal{L}_{SM}(\Psi) + \mathcal{L}_{SM}(\psi). \tag{5}$$

The $\Psi$ and $\psi$ parts differ only in weak interactions: *L* is the current for $\psi$ and *R* is the current for $\Psi$.

The Lagrangian that forms neutrino masses is assumed to have the following form [7]:

$$A\bar{\Psi}_{LR}\Psi_{LR} + B^{(\nu)}\bar{\Psi}_{RL}\Psi_{RL} + h^{(\nu)}\bar{\Psi}_{LR}\Psi_{RL}\varphi_1 + h^{(\nu)}\bar{\Psi}_{LR}\gamma_5\Psi_{RL}\varphi_2 - \\ - h_M \Psi_{RL}^T C \Psi_{RL}\varphi' - h_M \Psi_{RL}^T C \gamma_5 \Psi_{RL}\varphi'' + \text{c.c.} \quad (6)$$

where $A$ and $B^{(\nu)}$ are mass matrices (in the space of generation indices) of the isodoublet $\Psi_{LR}$ and isoscalar $\Psi_{RL}$. In the general case, these are Hermitian matrices. The weak symmetry *SU*(2) requires that $A$ be equal for the $\bar{u}$ and $\bar{d}$ components of the isotopic doublet; consequently, for neutrinos and charged leptons:[1]

$$A^{(\ell)} = A^{(\nu)} \equiv A, \quad B^{(\ell)} \neq B^{(\nu)}. \quad (7)$$

The bosons $\varphi_1, \varphi_2$ are isospinors and $\varphi', \varphi''$ are isoscalars.

In the most general case, the interaction matrices $h^{(\nu)}$ can be arbitrary, complex, while $h_M$ are complex, symmetrical. They are similar for each scalar-pseudoscalar pair. Such is the requirement of MS. If $h$ are equal, it will be impossible to determine physically whether the coordinate system being used is left-handed (*L*) or right-handed (*R*). The possibility of determining this was considered to be the main paradox of direct parity nonconservation [14,8]. In the proposed MS model, MS violation produces systems with small masses of $\psi$ or small masses of $\Psi$. If $h$ are equal, it is impossible to identify in which of these two states the system is, and without this, the *L*, *R* character of the coordinate system cannot be determined physically.

In terms of $\Psi$ and $\psi$, Eq.(6) is written out as follows [8]:

$$A\left(\bar{\Psi}_R \psi_L + \bar{\psi}_L \Psi_R\right) + B^{(\nu)}\left(\bar{\Psi}_L \psi_R + \bar{\psi}_R \Psi_L\right) + h^{(\nu)}\bar{\Psi}_R \psi_L \Phi_1 + h^{(\nu)}\bar{\psi}_L \psi_R \Phi_2 - \\ - h_M \Psi_L^T C \Psi_L \Phi' - h_M \psi_R^T C \psi_R \Phi'' + \text{c.c.}, \quad (8) \\ \Phi_{1,2} = \varphi_1 \mp \varphi_2, \quad \Phi^{',''} = \varphi' \mp \varphi''.$$

Using unitary transformations in the space of generation indices ($a, b$), it is possible, without loss of generality, to present $A$ and $B$ as diagonal real matrices, i.e., $\Psi_{LR}$ and $\Psi_{RL}$ masses, even in the initial Lagrangian. In all such transformations, the ratio (7) is of course not affected. An investigation of WMM properties, including complexity, is easier to start with the ($A, B$) diagonal form.

For Dirac particles—that is, charged leptons and quarks—masses are formed, in [6-9], by the Lagrangians of the type (6), (8), which, however, lack Majorana terms and use their own constants $h^{(l)}$, $B^{(l)}$ and so on. The isospinors $\Phi_1$ and $\Phi_2$ are the same for all types of particles (they produce *W* masses and the Higgs boson *H*). Formation of condensates [7] with vacuum averages

---

[1] This very equality defines the properties of the CKM matrix [6].

$$\langle \Phi_1 \rangle \;=\; \eta \quad \text{or} \quad \langle \Phi_2 \rangle \;=\; \eta \tag{9}$$

defines masses of weak $W_\mu$ bosons and masses of mirror fermions in the two possible "worlds" of violated MS—heavy $\Psi$ and heavy $\psi$. The emergence of the condensates $\langle \Phi_1 \rangle$ and $\langle \Phi' \rangle$ will leave $\Phi_2$ and $\Phi''$ as heavy boson states [7]. Therefore, the mass terms of the Lagrangian (8) for neutrinos can be written out as follows:

$$A\left(\bar{\Psi}_R \psi_L + \bar{\psi}_L \Psi_R\right) + B^{(\nu)}\left(\bar{\Psi}_L \psi_R + \bar{\psi}_R \Psi_L\right) + \mu^{(\nu)} \bar{\Psi}_R \Psi_L - \mathcal{M}_L \Psi_L^T C \Psi_L + \text{c.c.}$$
$$\mu^{(\nu)} = h^{(\nu)} \eta, \quad \mathcal{M} = h_M \langle \Phi' \rangle. \tag{10}$$

Compliance with phenomenology in [6-9] takes place if the ratio between the eigenvalues of the matrices in (10)

$$A \sim B \;\ll\; \mu^{(\nu)} \;\ll\; M, \tag{11}$$

is analogous to the see-saw conditions [11].

At arbitrary parameters, Eq.(10) results in $\psi$-Majorana states. Although masses of Majorana states can be small under condition (11), the neutrino spectrum, in contrast to Dirac, does not show potential for inverse character and the WMM based on these states does not demonstrate the observed qualitative properties. As a result, numerical selection of multiple parameters is required to achieve the reproduction.

Eq.(10) has another disadvantage which makes it unattractive for the MS approach. In addition to the weak interaction and the difference between masses of SM states and mirror fermions, it contains one more mechanism for parity violation. In fact, it includes only one chirality $\Psi_L$ (isoscalar) in the Majorana term. Dirac neutrinos cannot appear unless the isodoublet $\Psi_R$ is also present. As is well-known [13], obtaining the Majorana part of the mass Lagrangian with an isodoublet is a complicated procedure (8)-(10). For such a term to appear spontaneously, an isovector scalar or non-renormalized terms with the isodoublets $\Phi_1(\Phi_2)$ squared would have to be used. Neither option appears attractive.

In [7], a qualitative mechanism is proposed whereby, under MS, the term $\Psi_R$ appears under dynamics created as a result of *SU*(2) violation by the condensate (9). Let us assume that the Majorana part necessary for phenomenology

$$\mathcal{M}_R \Psi_R^T C \Psi_R \;+\; \text{c.c.} \tag{12}$$

is, in this or that way, included in the mass Lagrangian (10).

Parity conservation (operator $P = \gamma_0, R \leftrightarrow L$) and Dirac-type states co-exist in (10) and (12) only if

$$\mathcal{M}_R = -\mathcal{M}_L = \mathcal{M}. \tag{13}$$

($\mathcal{M}_L = \mathcal{M}_R$ is the Majorana system, see Appendix in [8]). If the *L*-Majorana mass term is presented in the Lagrangian, the dynamical mechanism [7] yields an obligatory appearance of the *R*-term in just the same form (after MS violation). This mechanism supports the possibility of opposite signs for both terms, but the equality of their modules (13) is an external requirement.

The Dirac nature also requires that the isodoublet and isoscalar masses of neutrinos in the MS Lagrangians (6), (8) be equal:[2]

$$A = B^{(\nu)}. \tag{14}$$

Equality (14) must be fulfilled not only for diagonal, but also for non-diagonal (in other representations of generations) mass matrices $A$ and $B$. This means that the Yukawa couplings $h^{(\nu)}$ in (8) and, consequently, the mass matrices $\mu^{(\nu)}$ in (10) must be Hermitian—that is, the transformation matrices $\Psi_R$ and $\Psi_L$ are the same for neutrino states.

Note that by selecting Dirac conditions—(13), (14), Hermitian $h$—we obtain a neutrino system in which parity is violated only by weak interactions. All other parts of the Lagrangian including neutrino terms conserve parity even after MS violation. This characteristic, so conceivable and so natural for the mirror symmetry approach, is also one of the conditions for the appearance, in the proposed model, of the qualitative properties of the PMNS matrix.

## 3. Neutrino States and Mass Matrices in MS See-Saw Model

An MS system selected using conditions in Section 2 allows the sum of the Lagrangians (10) and (12) for neutrinos to be written out in terms of full Dirac spinors $\Psi$ and $\psi$:

$$\begin{aligned} A_d(\bar{\psi}\Psi + \bar{\Psi}\psi) + \mu\bar{\Psi}\Psi - \mathcal{M}\Psi^T C\gamma_5\Psi + \mathcal{M}^+\bar{\Psi}\gamma_5 C\bar{\Psi}^T, \\ \Psi = \Psi_L + \Psi_R, \quad \psi = \psi_L + \psi_R. \end{aligned} \tag{15}$$

Here we use the real representation $C = i\gamma_0\gamma_2$, $C^+ = C^T = -C$. In the most general form, $A_d$ is a diagonal real matrix; $\mu$ is a Hermitian matrix; and $\mathcal{M}$ is a complex symmetrical matrix (3 x 3). Eq.(15) conserves parity: $\Psi \rightarrow \gamma_0\Psi$, $\psi \rightarrow \gamma_0\psi$.

---

[2] Condition (14) is missing in the initial version of [8]. Corrections were subsequently made in the text of the arXive publication and included as an addendum in Yad. Fiz. (Phys. Atom. Nucl.) **81**, 406 (2018).

The matrix $\mathcal{M}$ is diagonalized by a matrix satisfying the condition $U_M^T U_M = 1$. This can be seen from comparing the number of arbitrary parameters in the left-right sides of the equality:

$$\mathcal{M} = U_M \left\{ \frac{1}{2} M_d e^{i\delta_d} \right\} U_M^T, \quad U_M^T U_M = 1, \tag{16}$$

where $\{1/2\, M_d e^{i\delta_d}\}$ is a diagonal matrix of Majorana masses and phases [11]. The Hermitian matrix $\mu$ is diagonalized by unitary matrices $U$

$$U^+ U = 1, \quad \mu = U\{\mu_d\}U^+ \tag{17}$$

resulting in the diagonal real matrix $\{\mu_d\}$.

It can be seen that the Dirac system is produced only if the diagonalization of $\mu$ and $M$ occurs independently and simultaneously, i.e., via transformation of $U\Psi$ with the same matrix $U$. This means that

$$U = U_M, \quad U_M^T = U^+. \tag{18}$$

Thus $U$ must be a real orthogonal matrix. In other cases, pairs with equal masses—a sign of potential Dirac nature—are absent.

The matrix $U$ must also participate in the diagonalization of the Lagrangian for charged leptons, i.e., be a transformation for the whole isodoublet $U\Psi_{LR}$. This is necessary so that $A$, in any representation, fulfills equality (7). Only in this case, the properties of the PMNS matrix (as well as the CKM matrix for quark parameters) can be reproduced directly, without cumbersome, additional fitting of multiple constants [9]. At that, the matrix

$$A = U A_d U^T \tag{19}$$

remains real.

The real matrix $U$ makes the presence of Majorana phases in (16) unlikely. However, even if they are present, the transfer of the phases $\Psi' = \exp(i\delta/2)\Psi$ onto the matrix $A$ for neutrinos

$$A^{(\nu)} = U A_d U^T \{e^{-i\delta/2}\} = A\{e^{-i\delta/2}\} \tag{20}$$

does not affect the mass matrix of $\psi$ states and, consequently, the PMNS matrix (see Eq.(30)). In this case, the Majorana phases (16) are present in $\Phi''$ interactions, in weak charged currents of heavy mirror particles and their decay interactions.

The Lagrangian (15) with the diagonalized $\mu$ and $M$:

$$A^{(\nu)}\bar{\psi}\Psi + A^{(\nu)+}\bar{\Psi}\psi + \sum_{n=0}^{2}\left\{\mu_n\bar{\Psi}_n\Psi_n - \frac{1}{2}M_n\big(\bar{\Psi}_n^{(c)}\gamma_5\Psi_n - \bar{\Psi}_n\gamma_5\Psi_n^{(c)}\big)\right\}, \quad (21)$$
$$\Psi^{(c)} = C\bar{\Psi}^T, \quad \bar{\Psi}^{(c)} = \Psi^T C,$$

determines the states and masses of both mirror neutrinos and SM particles. The numbers $n = 0,1,2$ (as in [6-9]) for heavy mirror particles make the generation indices of these particles distinct from the generation indices of the $\psi$ states ($a, b = 1,2,3$).

This problem can be solved using expansion in inequalities (11). $A$ being neglected, masses and states of heavy mirror neutrinos are determined by the matrix:

$$\begin{array}{cc} \bar{\Psi} & \bar{\Psi}^{(c)}\gamma_5 \end{array}$$
$$\begin{vmatrix} M/2 & \mu/2 \\ \mu/2 & -M/2 \end{vmatrix} \begin{array}{c} \gamma_5\Psi^{(c)} \\ \Psi \end{array}. \quad (22)$$

In (22), all $\Psi$'s indices are omitted; three matrices are assumed for $n = 0,1,2$.

The orthogonal matrix diagonalizing (22) is equal to:

$$U = \frac{1}{N}\begin{vmatrix} 1 & -\dfrac{\mu}{M+\lambda} \\ \dfrac{\mu}{M+\lambda} & 1 \end{vmatrix}, \quad N = \left[\frac{2\lambda}{M+\lambda}\right]^{1/2}, \quad \lambda = (M^2 + \mu^2)^{1/2}. \quad (23)$$

The eigenvalues and operator eigenfunctions (22) are equal to:

$$\begin{aligned}\lambda_1 &= 1/2\,\lambda, & \Psi_1 &= \frac{1}{N}\Big(\gamma_5\Psi^{(c)} + \frac{\mu}{M+\lambda}\,\Psi\Big); \\ \lambda_2 &= -1/2\,\lambda, & \Psi_2 &= \frac{1}{N}\Big(\Psi - \frac{\mu}{M+\lambda}\,\gamma_5\Psi^{(c)}\Big).\end{aligned} \quad (24)$$

The eigenfunctions have the following property

$$\Psi_1^{(c)} = -\gamma_5\Psi_2, \quad \bar{\Psi}_1^{(c)} = \bar{\Psi}_2\gamma_5, \quad (25)$$

which leads to the following formula:

$$\bar{\Psi}_1^{(c)}\Psi_1^{(c)} = \bar{\Psi}_1\Psi_1 = -\bar{\Psi}_2\Psi_2. \quad (26)$$

Consequently, the $\Psi$ part of the Lagrangian is equal to:

$$\mu \bar{\Psi}\Psi - \frac{M}{2}\bar{\Psi}^{(c)}\gamma_5 \Psi + \frac{M}{2}\bar{\Psi}\gamma_5 \Psi^{(c)} = \frac{\lambda}{2}\big(\bar{\Psi}_2 \Psi_1 + \bar{\Psi}_1 \Psi_2\big) =$$
$$= \lambda \frac{(\bar{\Psi}_1 + \bar{\Psi}_2)}{\sqrt{2}} \frac{(\Psi_1 + \Psi_2)}{\sqrt{2}} \equiv \lambda \bar{\Psi}_\lambda \Psi_\lambda, \quad \Psi_\lambda = \frac{\Psi_1 + \Psi_2}{\sqrt{2}}. \tag{27}$$

Eq.(27) is a Dirac particle $\Psi_\lambda$ with the mass $\lambda$.

Let us now consider the character of the $\psi$ states that result from the Lagrangian (21). In Appendix 1, it is shown that the kinetic part of the general Lagrangian (5), both in terms of $\Psi$ and in terms of $\Psi_\lambda$, has a similar form. $\Psi_\lambda$ represents a truly Dirac state. Then the mass matrix $\psi$ can be found from self-energy diagrams. At $m_\psi \ll \lambda$ (assuming $m_\psi \sim m_{SM}$), $|\hat{p}| \approx m_\psi$ in the propagator $\Psi$ can be neglected. From the formula

$$(\bar{\psi} m_\nu \psi) = \bar{\psi}_\alpha A^{(\nu)} \langle \Psi_\alpha, \bar{\Psi}_\beta \rangle A^{(\nu)+} \psi_\beta \tag{28}$$

($\alpha, \beta$ are spinor indices), with the operators $\Psi$ expressed in $\Psi_\lambda$:

$$\Psi = \frac{1}{\sqrt{2}\,N}\bigg[\Big(1 + \frac{\mu}{M+\lambda}\Big)\Psi_\lambda - \gamma_5\Big(1 - \frac{\mu}{M+\lambda}\Big)\Psi_\lambda^{(c)}\bigg],$$
$$\bar{\Psi} = \frac{1}{\sqrt{2}\,N}\bigg[\Big(1 + \frac{\mu}{M+\lambda}\Big)\bar{\Psi}_\lambda + \Big(1 - \frac{\mu}{M+\lambda}\Big)\bar{\Psi}_\lambda^{(c)}\gamma_5\bigg] \tag{29}$$

and taking into consideration that only the $\langle \Psi_\lambda \bar{\Psi}_\lambda \rangle$ propagators are different from zero, we arrive at the following formula for the Dirac part of the neutrino mass operator $\psi$:

$$(m_\nu^{(D)})_{ab} = \frac{1}{2N^2}\sum_{n=0}^{2} A_{an}^{(\nu)}\bigg[\Big(1 + \frac{\mu}{M+\lambda}\Big)^2 - \gamma_5\Big(1 - \frac{\mu}{M+\lambda}\Big)^2 \gamma_5\bigg]_n \times$$
$$\times \frac{1}{\lambda_n} A_{nb}^{(\nu)+} = \frac{1}{N^2}\sum_{n=0}^{2}\Big(A\,\frac{2\mu}{(M+\lambda)\lambda}\,A\Big)_{ab}. \tag{30}$$

Note that the Majorana phase $\delta$ has disappeared from this expression. The Dirac part of the neutrino mass matrix is real (in the $|A| \ll M$ approximation).

A similar calculation for the Majorana part of the mass matrix $\psi$ ($\alpha, \beta$ are spinor indices):

$$\sum_{n=0}^{2} \psi_\alpha^T A^{(\nu)T} \langle \bar{\Psi}_\alpha, \bar{\Psi}_\beta \rangle A^{(\nu)} \psi_\beta,$$
$$\langle \bar{\Psi}_\alpha, \bar{\Psi}_\beta \rangle = \delta_{\alpha,\beta}\bigg[1 - \Big(\frac{M}{M+\lambda}\Big)^2\bigg]\frac{1}{\lambda}\bigg[-(C\gamma_5)^T_{\beta\alpha} + (C\gamma_5)_{\beta\alpha}\bigg] \equiv 0, \tag{31}$$

also indicates the Dirac character of $\psi$. These are neutrinos of the system (21) with very tiny masses (30).

## 4. Complex Parameters of PMNS Matrix

In the lowest order of the parameters (11) under consideration, the only possible complexity of the neutrino Lagrangian (15), (21)—the Majorana phase $\delta$—is not transferred, during diagonalization, to the wavefunctions of physical neutrinos. CP-violating phases can appear only from charged leptons.

In the charged system, in the $A/\mu \ll 1$ approximation, we obtain, upon diagonalization of the respective couplings, the mass matrix [6, 8, 9]:

$$(m^{(\ell)})_{ab} = \sum_{n=0}^{2} A_{an}^{(\ell)} \frac{1}{\mu_n^{(\ell)}} B_{nb}^{(\ell)+} . \tag{32}$$

The three-dimensional vectors $A_n^\ell, B_n^\ell, n = 0,1,2$ with projections on the axes of generation indices $a, b = 1,2,3$ demonstrate here the following properties (in contrast to conditions (14) for neutrinos):

$$A^{(\ell)} \neq B^{(\ell)}, \quad \text{but} \quad A^{(\ell)} = A^{(\nu)} = A . \tag{33}$$

As explained earlier, the vector $A$, which is common for the systems $\ell$ and $\nu$, is a real quantity. Only vectors $B^\ell$ can be the source of complexity in the PMNS matrix.

The PMNS matrix elements are scalar products (in the space of generation indices) of the wavefunctions of the physical states $\phi_\ell$ and $\phi_\nu$, which represent charged leptons and neutrinos of various massive generations.

In [6], diagonalization of a separable matrix similar to Eq.(30) or (32) is completed for such a case where a spectrum of physical states must obey a resultant hierarchical structure. It is assumed that such a structure is provided by the hierarchy of the Yukawa masses $\mu_n$:

$$m_\tau \gg m_\mu \gg m_e \longleftrightarrow \mu_2 \gg \mu_1 \gg \mu_0 . \tag{34}$$

Vectors $A$ and $B$ (33) have, more naturally, close "lengths" (even if masses $A_d$, $B_d$ are essentially different). Under the inverse hierarchy in Eq.(32), the heaviest mass $m_r$ matches the lightest mass $m_0$ and so on. The same hierarchy $\mu_i$, which is present in the numerators of Eq.(30), suggests that the SM-neutrino spectrum is of the inverse type [7], which essentially facilitates reproduction of "peculiar" qualitative properties of the PMNS matrix.

Results [6-9] show that the complex vectors $B^\ell$ are present in the wavefunctions of charged leptons only in the correction terms of hierarchy (34). The principal contributions to these functions depend only on real vectors $A$. As a result, mass complexities are accompanied by small mass ratio factors for leptons of different generations:

$$\frac{m_e}{m_\mu} \approx 0.005 \quad \text{or} \quad \frac{m_\mu}{m_\tau} \approx 0.06. \tag{35}$$

Investigation into WMM complexity cannot be complete without clarifying one other complicated matter. Parametrization of the mass matrices (30) and (32) does not include the canonical parameters of lepton WMM—the mixing angles of neutrino generations $\theta_{12}$, $\theta_{13}$, $\theta_{23}$. MWWs calculated from (30), (32) can therefore differ from the standard form [11] in wavefunction phase transformations. This requires that the WMM be reduced to the PMNS matrix.

These questions are discussed in Appendix 2. The appendix involves cumbersome calculations and draws heavily on [6, 8, 9], being based on the properties discovered in these papers. For the standard form of a PMNS element (neutrino3–electron)—$V_{3e} = sin\theta_{13}\, expi\delta_{CP}$—we obtain (A18):

$$V_{3e} \simeq (0.14-0.16) + \xi F(b_n)\frac{m_e}{m_\mu},$$

where $|\xi| \sim 1$, $|F(b_n)| \lesssim 1$ is a complex (approximately purely imaginary) factor. To derive Eq.(A18), we will use both the theoretical conclusions from [6-9] and experimental data [15] that allow evaluation of the parameters used in this paper. Eq.(A18) proves that the complex phase $\delta_{CP}$ in the proposed MS scenario provides a small $|sin\delta_{CP}|$.

## 5. Conclusion

Masses $m_\nu$ appearing during diagonalization (30) are parametrically more indicative of a larger difference between neutrino and charged lepton masses than is suggested by the conventional see-saw formula [13]. We obtain:

$$m_\nu \sim m^{(\ell)}\frac{\mu}{M} \text{ (see-saw)}, \quad m_\nu \sim m^{(\ell)}\left(\frac{\mu}{M}\right)^2 \text{(MS)} \tag{36}$$

In (36), we assume that for see-saw $\mu \sim m^\ell$ are charged lepton masses, and for MS, $m^\ell \sim AB/\mu$ [8]. In both cases, we assume that the parameters $\mu^\ell \sim \mu^\nu$. This is also assumed in the standard see-saw scenario.

Eq.(A18) demonstrates that the CP-violating lepton phase $\delta_{CP}$ leads to the small $|sin\delta_{CP}|$. The ratio $m_e/m_\mu$ is significantly smaller than the observed real term in Eq.(A18). This term cannot be evaluated theoretically due to contributions from the neutrino mass ratios $m_3/m_1$, $m_3/m_2$, which are hard to calculate, hence neglected in the proposed model, and which are small in the case of inverse neutrino spectrum: $m \ll m_1 \sim m_2$. For the conditions being discussed, inclusion of such corrections leaves quantities real. In (A18), instead of calculating this contribution, we use the experimental value [11].

Majorana phases, lepton number nonconservation (see Eq.(A7) in Appendix 1), and $CP = T$ invariance violation [16] are fully manifest only in processes involving heavy neutral neutrinos.

Subsequent approximations of the small parameter $m_{SM}/M_{mir}$ may impose additional requirements for the preservation of the Dirac character of SM neutrino. These would be new theoretical conditions for all parameters, including, ultimately, lepton masses and WMM elements.

The proposed MS violation mechanism (via vacuum averages of scalar fields) leads to a nonperturbative coupling of the Higgs boson $H$ ($m_H = 126$ GeV) with very heavy mirror fermions: $M_{mir} \gg m_H$. The effect of non-perturbativity is therefore observed in phenomena over exceptionally small distances ($1/M_{mir}$). Their contributions to the processes involving SM energies and particles appear to be insignificant. This is considered in [17] where heavy fermion contributions from small distances to the production cross-section of the Higgs boson are assessed.

## Appendix 1

Let us show that the transition from fermions $\Psi$ to representatives of massive neutrinos $\Psi_\lambda$ preserves the form of the kinetic part. In weak currents of heavy mirror neutrinos and in their interactions with the Higgs boson, such a transition produces terms violating lepton number conservation. This fact has already been shown in [16].

Let us consider the neutral current $\Psi$. By substituting Eq.(29) we obtain:

$$\bar{\Psi}\gamma_\mu\Psi = \frac{1}{2N^2}\Big\{\Big(1+\frac{\mu}{M+\lambda}\Big)^2 \bar{\Psi}_\lambda \gamma_\mu \Psi_\lambda + \Big(1-\frac{\mu}{M+\lambda}\Big)^2 \bar{\Psi}_\lambda^{(c)} \gamma_\mu \Psi_\lambda^{(c)}\Big\} +$$
$$+ \frac{1}{2N^2}\Big(1-\frac{\mu^2}{(M+\lambda)^2}\Big)\Big\{-\bar{\Psi}_\lambda \gamma_\mu \gamma_5 \Psi_\lambda^{(c)} + \bar{\Psi}_\lambda^{(c)} \gamma_5 \gamma_\mu \Psi_\lambda\Big\}. \tag{A1}$$

Both brackets are not equal to zero:

$$\bar{\Psi}_\lambda^{(c)} \gamma_\mu \Psi_\lambda^{(c)} = \Psi_\lambda^T C \gamma_\mu C \bar{\Psi}_\lambda^T = -\bar{\Psi}_\lambda \gamma_\mu \Psi_\lambda, \tag{A2}$$

since $\Psi$ anticommute, $C^2 = -1$. In the second bracket

$$\bar{\Psi}_\lambda \gamma_\mu \gamma_5 \Psi_\lambda^{(c)} = \bar{\Psi}_{\lambda L} \gamma_\mu \gamma_5 C \bar{\Psi}_{\lambda R}^T + \bar{\Psi}_{\lambda R} \gamma_\mu \gamma_5 C \bar{\Psi}_{\lambda L}^T, \tag{A3}$$

and the second term in Eq.(A3) is equal to the first one:

$$\bar{\Psi}_{\lambda R}\gamma_\mu\gamma_5 C\bar{\Psi}^T_{\lambda L} = \bar{\Psi}_{\lambda L}\gamma_\mu\gamma_5 C\bar{\Psi}^T_{\lambda R}. \tag{A4}$$

In the kinetic term of the Lagrangian, the derivatives $\hat{p} = i\gamma\partial$ in the second term of the first bracket in Eq.(A1), i.e., Eq.(A2), and in the second term (A3), i.e., Eq.(A4), should be redirected from one $\Psi$ to the other. The minus sign resulting from this operation will make the contribution from the second bracket equal to zero; the first bracket, according to (23), is equal to:

$$\bar{\Psi}\hat{p}\Psi = \frac{1}{2N^2}\left\{\left(1 + \frac{\mu}{M+\lambda}\right)^2 + \left(1 - \frac{\mu}{M+\lambda}\right)^2\right\}\bar{\Psi}_\lambda \hat{p}\Psi_\lambda \equiv \bar{\Psi}_\lambda \hat{p}\Psi_\lambda \tag{A5}$$

—that is, the kinetic part of the fermion $\Psi_\lambda$.

By calculating the weak neutral current of heavy mirror neutrinos $\Psi$

$$\bar{\Psi}^{(\nu)}_R \gamma_\mu \Psi^{(\nu)}_R = \bar{\Psi}^{(\nu)}\gamma_\mu \frac{1+\gamma_5}{2}\Psi^{(\nu)} \tag{A6}$$

we obtain, again using Eq.(29), the expression which does not conserve lepton numbers:

$$\frac{1}{2N^2}\left\{\bar{\Psi}_\lambda\left(1 - \frac{\mu}{M+\lambda}\gamma_5\right)^2\gamma_\mu\gamma_5\Psi_\lambda - \left(1 - \frac{\mu^2}{(M+\lambda)^2}\right) \times \right.$$
$$\left. \times \left[\bar{\Psi}^{(c)}_\lambda \gamma_\mu \frac{1+\gamma_5}{2}\Psi_\lambda + \bar{\Psi}_\lambda\gamma_\mu\frac{1+\gamma_5}{2}\Psi^{(c)}_\lambda\right]\right\}. \tag{A7}$$

Eq.(A7) is analogous to Eq.(39) in [16], which was calculated in a different way. Weak charged currents and mirror neutrino interactions with scalars are calculated in a similar fashion.

In the higher orders of the small parameter $m_{SM}/M_{mir} \ll 1$, lepton number nonconservation may also appear in the system of SM light neutrinos in the form of minor corrections to SM formulae.

## Appendix 2

In [6], we found the wavefunctions (in the space of generation indices) of massive charged fermions of SM. These wavefunctions are eigenfunctions of the separable matrix (32)—the mass matrix of the MS model. Calculations are carried out using the perturbation theory for hierarchy (34).

For the vectors $A_n$ and $B_n$ [6], let us apply parametrization as per this paper:

$$A_n \Rightarrow A_n a_n, \quad B_n \Rightarrow \frac{B_n}{\mu_n} b_n, \quad n = 0, 1, 2. \tag{A8}$$

where $A_n$ and $B_n$ in the right part of Eq.(A8) are vector moduli, normalized vectors: $a_n = (a_1^{(n)}, a_2^{(n)}, a_3^{(n)})$ and similarly $b_n$: ($|b_n| = 1$ − complex and $|a_n| = 1$ − real). Leaving only the complex contributions that are the largest in hierarchy (34), we obtain the left-handed wavefunctions of charged leptons (Eqs.(28-30), [6]):

$$\Phi_\tau = a_0 + \frac{(B_1 A_1/\mu_1)}{(B_0 A_0/\mu_0)}(b_1^+, b_0)a_1 + \cdots,$$

$$\Phi_\mu = \frac{a_1 - \cos\alpha_{01} a_0}{|\sin\alpha_{01}|} - |\sin\alpha_{01}|\frac{(B_1 A_1/\mu_1)}{(B_0 A_0/\mu_0)}(b_0^+, b_1)a_0 + \cdots, \quad (A9)$$

$$\Phi_e = \frac{[a_0, a_1]}{|\sin\alpha_{01}|} - F(b_n)\frac{(B_2 A_2/\mu_2)}{(B_1 A_1/\mu_1)}\frac{[a_0, a_2]}{|\sin\alpha_{01}|} + \cdots,$$

Where the complex factor $F(b_n)$ is equal to

$$F(b_n) = \frac{(b_1^+, b_2) - (b_1^+, b_0)(b_0^+, b_2)}{1 - |(b_1^+, b_0)|^2}. \quad (A10)$$

Vector product modules of the vectors are

$$|[a_0, a_1]| = \sin\alpha_{01}, \quad |[a_0 a_2]| = \sin\alpha_{02} \text{ and so on}, \quad (A11)$$

where $\alpha_{01}$, $\alpha_{02}$, $\alpha_{12}$ are angles between real vectors $a_n$. The scalar products $a_n$ are: $(a_0, a_1) = \cos\alpha_{01}$ and so on.

The wavefunctions of the Dirac neutrino are real (Section 3). To calculate the largest possible complex contributions to the WMM, it is sufficient to use the principal terms of these functions. Based on Eq.(40), ([8]), if the neutrino spectrum were inverse and purely hierarchical—$m_1 > m_2 > m_3$—then the main terms of the neutrino left functions (the eigenfunctions of matrix (30)) would be:

$$\Phi_{\nu_1} = a_2 + \ldots,$$

$$\Phi_{\nu_2} = \frac{a_1 - \cos\alpha_{12} a_2}{|\sin\alpha_{12}|} + \ldots \quad (A12)$$

$$\Phi_{\nu_3} = \frac{[a_2, a_1]}{|\sin\alpha_{12}|} + \ldots.$$

In the actual observed neutrino spectrum, states with numbers 1,2 have closely situated masses ("degeneracy", [8]), which are the largest under the inverse character of the spectrum. The wavefunctions of the states 1,2 are then determined using Eq.(A12) and Eq.(38) from [9]:

$$\Phi'_1 = \Phi_{\nu_1} \cos\beta + \Phi_{\nu_2} \sin\beta ,$$
$$\Phi'_2 = -\Phi_{\nu_1} \sin\beta + \Phi_{\nu_2} \cos\beta .$$
(A13)

The real angle $\beta$—"degeneracy removal angle"—depends on corrections to functions (A12) and the influence of $\Phi_{\nu_3}$ on these functions. In the proposed model, $\beta$ remains one other parameter of the problem.

From Eq.(A9) and Eq.(A12), the element $V_{3e}$ of the lepton WMM is expressed as follows:

$$V_{3e} = \langle \Phi_{\nu_3}, \Phi_e \rangle = \frac{([a_2,a_1],[a_0,a_1])}{|\sin\alpha_{01}\sin\alpha_{12}|} - F(b_n)\frac{(B_2 A_2/\mu_2)}{(B_1 A_1/\mu_1)}\frac{([a_2,a_1],[a_0,a_2])}{|\sin\alpha_{01}\sin\alpha_{12}|} ,$$

$$([a_2,a_1],[a_0,a_1]) = \cos\alpha_{02} - \cos\alpha_{01}\cos\alpha_{12} ,$$
$$([a_2,a_1],[a_0,a_2]) = \cos\alpha_{02}\cos\alpha_{12} - \cos\alpha_{01} .$$
(A14)

The first (real) term in Eq.(A14) appears to be a small quantity [9]. In fact, for the hierarchical character of the charged lepton mass spectrum, the vector $a_2$ is approximately orthogonal to vectors $a_0$ and $a_1$ with a precision to small mass ratios of different generations. In the MS approach, approximate orthogonality provides theoretical justification for the smallness of the neutrino mixing angle $\theta_{13}$ (Eq.(35), Eq.(37), [9]). However, the first term in Eq.(A14) cannot determine on its own the value of $V_{3e}$. Additional real contributions of the same order of magnitude can be provided by the inclusion of the small mass $m_3$ and Eq.(12) and Eq.(A13) corrections in these calculations. The sum of all these quantities must define the absolute value of the element $|V_{3e}| = |\sin\theta_{13}| \approx 0.14\text{-}0.16$ if the complex contribution to (A14) is even smaller.

The parameters defining the lepton WMM in Eq.(A.14) differ from the standard mixing angles $\theta_{12}, \theta_{13}, \theta_{23}$ of neutrino generations. Therefore, the complex factor in (A14) can be changed by transferring complex phases of other WMM elements by means of redetermination of lepton wavefunction phases. Such phases can be transferred to the element $V_{3e}$ from the elements $V_{3r}$ and $V_{1e}$. In the approximation being considered, the element $V_{3r}$ proves to be real and the element $V_{1e}$ is equal to

$$V_{1e} = \langle \phi'_{\nu_1} | \phi_e \rangle = \frac{(a_2,[a_0,a_1])}{|\sin\alpha_{01}|} \times$$
$$\times \left[ (\cos\beta - \sin\beta |\cot\alpha_{12}|) + F(b_n)\frac{(B_2 A_2/\mu_2)}{(B_1 A_1/\mu_1)}\frac{\sin\beta}{|\sin\alpha_{12}|} \right],$$
(A15)

i.e., the complex factor is the same as in Eq.(A14). This factor is unconditionally small in the proposed MS system and is related with the mass ratio $m_e/m_\mu$. In fact, as per Eq.(24) and Eq.(25) in [6], inclusion of (A8) results in:

$$m_e \simeq \frac{A_2 B_2}{\mu_2} f_e(a_n, b_n), \quad m_\mu \simeq \frac{A_1 B_1}{\mu_1} f_\mu(a_n, b_n), \tag{A16}$$

$f_e, f_\mu$ are real functions determined in [6].

Let us prove that all factors accompanying $F(b_n)$ complexities cannot be significantly larger than unity. For this purpose, we use the following properties:

1. The approximate vector orthogonality $a_2 \perp a_0, a_1$ follows from the hierarchical character of the lepton spectrum [9].
2. The separable form of the mass matrix (32) leads to a similar approximate orthogonality $b_2 \perp b_0, b_1$ for purely real $b_n$.
3. Compared to the approximate skeleton of the lepton WMM built in [9] (see Eq.(39) in [9]), the observed PMNS matrix [15] allows evaluation of a number of parameters being used:

$$\cos\beta \simeq 0.8\text{--}0.9, \ \sin\beta \simeq 0.5\text{--}0.6, \ \sin\alpha_{12} \sim \cos\alpha_{12} \approx 0.6\text{--}0.8. \tag{A17}$$

As a result, we obtain the following expression for the standard form of the element $V_{3e}$:

$$V_{3e} \simeq (0.14\text{--}0.16) + \xi F(b_n) \frac{m_e}{m_\mu}, \tag{A18}$$

where $|\xi| \sim 1$. From Eq.(A10) it follows that $F(b_n)$ is a small (if $b_0 \neq b_1$), almost purely imaginary quantity, since at real $b_2 \perp b_0, b_1$, $F(b_n) \simeq 0$. One can see no apparent reason for the equality of the "directions": $b_0 = b_1$.

## Bibliography

1. M. Fukugita and T. Yamagida, Phys. Lett. B **174**, 45 (1986).

2. C. Hagedorn, R.N., R.N. Mohapatra, E. Molinaro, C.C. Nishi, and S.T. Petcov, J. Mod. Phys. A **33**, 1842006 (2018), arXiv: 1711.02866 [hep-ph].

3. K. Abe *et al.* (T2K Collab.), Phys. Rev. D **91**, 072010 (2015); arXiv:1502,01550 [hep-ph].

4. J. Elevant and T. Schwetz, arXiv:1506,07685 [hep-ph];
   D. V. Forero and P. Huber, arXiv:1601.03736 [hep-ph].

5. S. Pascoli, CERN Courier **56**, 6, p. 34 (2016).

6. I. T. Dyatlov, Yad. Fiz. **77**, 775 (2014); [Phys. Atom. Nucl. **77**, 733 (2014)]; arXiv:1312.4339 [hep-ph].

7. I. T. Dyatlov, Yad. Fiz. **78**, 1015 (2015) [Phys. Atom. Nucl. **78**, 956 (2015)]; arXiv:1509.07280 [hep-ph].

8. I. T. Dyatlov, Yad. Fiz. 78, 522 (2015) (corrigendum: Yad. Fiz. **81**, 406 (2018)). [Phys. Atom. Nucl. 78, 485 (2015)];arXiv:1502.01501 [hep-ph].

9. I. T. Dyatlov, Yad. Fiz. **80**, 368 (2017) [Phys. Atom. Nucl. **80**, 679 (2017)]; arXiv.1703.00722 [hep-ph].

10. L. Wolfenstein, Phys. Rev. Lett. **51**, 1945 (1983).

11. M. Takabashi et al. (Particle Data Group), Phys. Rev. D **98**, 030001 (2018)

12. F. P. An *et al.* [Daja Bay Collab.], Phys. Rev. Lett. **108**, 171803 (2012);
    J. K. Ahn *et al.* [RENO Collab.], Phys. Rev. Lett. **108**, 191802 (2012);
    Y. Abe *et al.* [Double Chooz Collab.], Phys. Rev. D **86**, 052008 (2012).

13. R. N. Mohapatra and A. Yu. Smirnov, hep-ph/0603118;

14. T. D. Lee and C. N. Yang, Phys. Rev. **104**, 254 (1956).

15. J. Bergstrom, M.C. Gonzalez-Garcia, M. Maltoni, T. Schwetz, JHEP **1509**, 200 (2015); arXiv: 1507.04366 [hep-ph].

16. I. T. Dyatlov, Yad.Fiz. **80**, 151 (2017) [Phys Atom. Nucl. **80**, 275 (2017)]; arXiv:1611.05635 [hep-ph].

17. I. T. Dyatlov, Yad.Fiz. **80**, 253 (2017) [Phys. Atom. Nucl. **80**, 469 (2017)].